\documentclass[
 reprint,
 amsmath,amssymb,
 aps, physrev,
 twocolumn,superscriptaddress
]{revtex4-2}

\usepackage{graphicx}
\usepackage{dcolumn}
\usepackage{bm}
\usepackage{hyperref}
\usepackage{xcolor}
\usepackage{float}

\usepackage[
text={7in,9.5in},centering,
]{geometry}
\usepackage{lipsum}

\begin{document}

\preprint{APS/123-QED}

\title{
    \textbf{Programmable Dynamic Phase Control of a Quasiperiodic Optical Lattice} 
}
\author{Andrew O. Neely}
    \affiliation{Department of Physics, Yale University, New Haven, CT 06520, USA}
\author{Cedric C. Wilson}
    \affiliation{Department of Physics, Yale University, New Haven, CT 06520, USA}
    \affiliation{Yale Quantum Institute, Yale University, New Haven, CT 06520, USA}
\author{Ryan Everly}
    \affiliation{Department of Physics, Yale University, New Haven, CT 06520, USA}
\author{Yu Yao}
    \affiliation{Center for Computational Science and Engineering, Massachusetts Institute of Technology, Cambridge, MA 02139, USA}
\author{Raffaella F. Zanetti}
    \affiliation{Department of Physics, Yale University, New Haven, CT 06520, USA}
\author{Charles D. Brown}
    \email{charles.d.brown@yale.edu}
    \affiliation{Department of Physics, Yale University, New Haven, CT 06520, USA}
    \affiliation{Yale Quantum Institute, Yale University, New Haven, CT 06520, USA}
\date{\today}

\begin{abstract}
The quantum dynamics of quasiperiodic systems display a rich variety of physical behaviors due to the combination of rotational symmetry that is mathematically forbidden in periodic systems, and long-range order despite the lack of translation symmetry. New experimental probes into these dynamics with a quantum simulator, consisting of ultracold atoms in an optical lattice potential, will yield new insights into the physics of quasiperiodic systems. This potential is imbued with the flexibility, tunability, and purity of the individual laser beams that constitute it, allowing for exquisite control over a rich system. Programmable dynamic control over the lattice beam phases opens up an even richer space of achievable systems via Floquet engineering. We thus describe an experimental scheme for creating a programmable, dynamic, two-dimensional quasiperiodic optical lattice with heavily suppressed phase noise. We observe suppression of phase noise for frequency components up to 5 kHz, and report phase noise suppression of over 70 dB over the DC-60 Hz frequency band. We further demonstrate a phase modulation bandwidth of 350 kHz. This scheme allows for full translational and phasonic control of the lattice, including changes to the rotational symmetry of the potential, at speeds exceeding the lattice recoil velocity of lithium, which paves a path towards direct observation and control of quantum dynamics in quasicrystals.
\end{abstract}

\maketitle

\section{\label{sec:intro}Introduction}

Quasicrystals, which are defined by their long-range order despite their lack of spatial periodicity, have captivated researchers since their discovery \cite{shechtman_metallic_1984}. The strange combination of symmetries present in these materials generates exotic behaviors, including  fractal wavefunctions \cite{ghadimi_mean-field_2020,rolof_electronic_2013} and many-body localization \cite{sbroscia_observing_2020}. Recent results from Moir\'e materials, such as twisted bi-layer graphene \cite{yao_quasicrystalline_2018}, have led to quasicrystals becoming the subject of increasing scientific intrigue, while new theoretical tools have accelerated the study of the topological physics \cite{gottlob_origin_2025,gottlob_quasiperiodicity_2025,huang_quantum_2018,dareau_revealing_2017,bandres_topological_2016,matsuda_topological_2014,tran_topological_2015,duncan_topological_2020,burgess_quasicrystalline_2026}, quantum chaos \cite{singh_fibonacci_2015}, and superconductivity \cite{kamiya_discovery_2018,araujo_conventional_2019,sakai_exotic_2019,sakai_superconductivity_2017,uri_superconductivity_2023} generated in these systems.

Meanwhile, advancements in atomic and optical physics have made cold atoms in optical lattices an appealing medium for realizing a quantum simulator of a quasicrystal, as the inherent tunability and purity of optical lattices allow access to behaviors that are difficult to observe in the solid state \cite{tarruell_creating_2012, hou_superfluid-quasicrystal_2018,spurrier_semiclassical_2018,gross_quantum_2017,shimasaki_reversible_2024,grass_colloquium_2025}. 

Recently, an eight-fold rotationally symmetric optical lattice formed from four intersecting optical standing waves was created \cite{viebahn_matter-wave_2019} and has since been used to study disorder-induced localization \cite{sbroscia_observing_2020} and the transition from superfluid to Bose glass in quasiperiodic systems \cite{yu_observing_2024}. A complimentary approach towards generating a QC lattice that focuses on dynamic phase control of the beams will enable an entirely new class of quantum dynamics experiments.

\begin{figure}[H]
\centering
\includegraphics[width=0.48\textwidth]{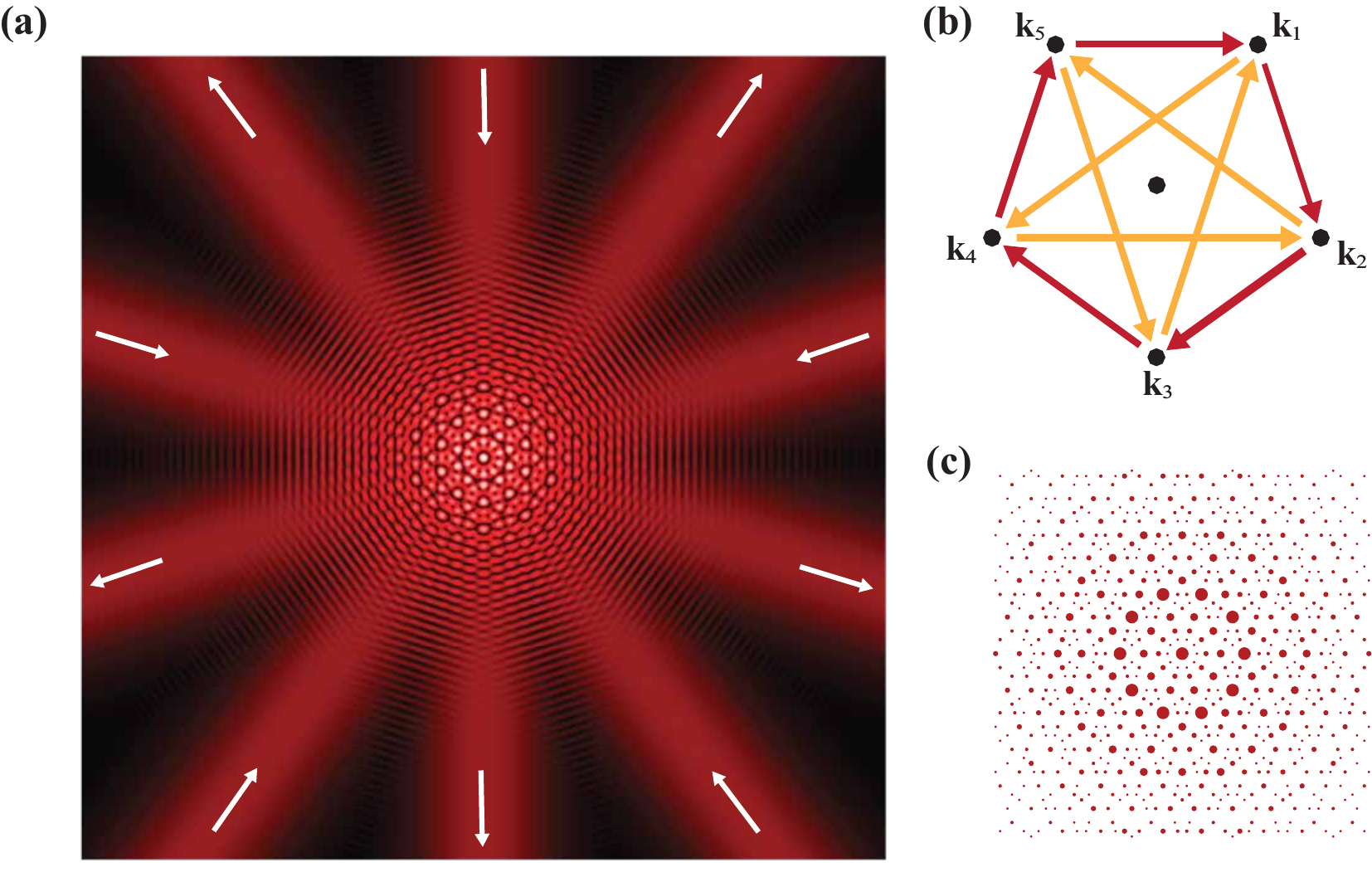}
\caption{\label{fig:fig1}An optical quasicrystal.  (a) Illustration of the optical quasicrystal. The mutual interference of five laser beams oriented at 72$^\circ$ forms the quasiperiodic lattice. (b) The five lattice beam wavevectors $\mathbf{k}_j$ define ten momentum difference vectors: five differences between nearest neighbor wavevectors (red), and five differences between next-nearest neighbor wavevectors (yellow). The optical quasicrystal is the superposition of ten standing waves, each having one of these ten difference vectors as its wavevector. (c) A simulated diffraction (momentum space) image of the quasiperiodic lattice. Integer combinations of wavevectors $\mathbf{k}_j$ reveal the 10-fold rotation symmetry of the quasicrystal.}
\end{figure}

Dynamic phase control in optical lattices has been demonstrated in periodic lattices \cite{brown_direct_2022,lang_edge_2012,kosch_multifrequency_2022,li_observation_2025,li_bloch_2016,mehling_high-precision_2026}, and has been utilized to great effect to study topological \cite{brown_direct_2022,li_bloch_2016,li_observation_2025} and geometric \cite{kosch_multifrequency_2022} behaviors of periodic systems. Dynamic control schemes for two-dimensional (2D) quasiperiodic lattices have been proposed \cite{kosch_multifrequency_2022,spurrier_semiclassical_2018,sanchez-palencia_bose-einstein_2005,corcovilos_two-dimensional_2019}, but not yet experimentally realized. 

Here, we describe a phase control scheme that provides control of a two-dimensional quasiperiodic optical lattice that arises from the mutual interference of five laser beams oriented at 72$^\circ$ to each other (see Fig. \ref{fig:fig1}). The phase stability demands of a 2D quasicrystal are greater than those of many other systems because phase errors can disturb the geometry of the quasicrystal. Since the lattice is formed by the interference of five mutually coherent traveling waves, it can be manipulated by applying phase shifts to the five beams. We accomplish this by making small adjustments to the optical frequency via adjustments of the drive tone of an acousto-optic modulator (AOM) on each of the five lattice beams. This scheme allows for the phases of the beams to be adjusted over arbitrary ranges at rates of up to 350 kHz, which is crucial to exploring the quantum dynamics of the quasicrystal. A similar dynamic lattice control technique has previously been demonstrated with a periodic optical lattice for experiments with ultracold atoms \cite{brown_direct_2022,zephy}. Independent dynamic control of the optical phases of these five laser beams allows for full dynamic control of the translational and phasonic degrees of freedom of the quasicrystal. This scheme opens a new class of experiments positioned to explore and directly probe the quantum dynamics of a quasicrystal, including quasiperiodic band structure and Bloch oscillations \cite{lu_electronic_1987,sanchez-palencia_bose-einstein_2005,spurrier_semiclassical_2018}, anomalous group velocity induced by Berry curvature \cite{spurrier_semiclassical_2018}, and Thouless pumping \cite{corcovilos_two-dimensional_2019,freedman_phason_2007,shimasaki_reversible_2024,kraus_topological_2012}.

\section{\label{sec:experiment}Apparatus}

\begin{figure}[!ht]
\includegraphics[width=0.45\textwidth]{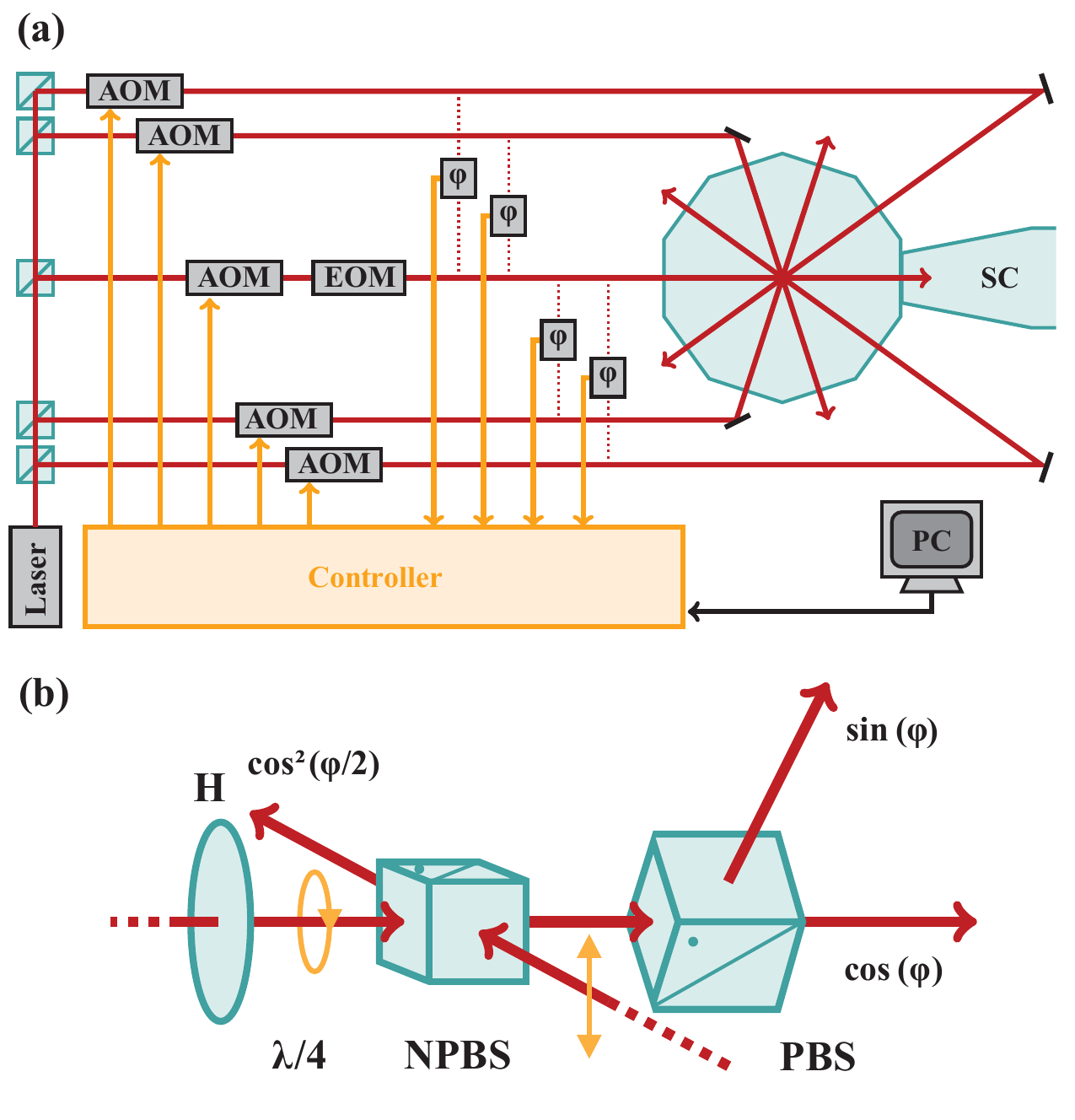}
\caption{\label{fig:fig2} Phase measurement and control system for the dynamic quasicrystal lattice. (a) Five beams from the same laser pass through acousto-optic modulator (AOMs). One beam (the phase reference) also passes through an electro-optic modulator (EOM). The beams' relative phases are measurede by phase detectors ($\varphi$) whose output is used to implement feedback control over the beams' phases. The five beams overlap inside the decagonal quartz science cell (SC) where they form a quasiperiodic lattice (Fig.~\ref{fig:fig1}). The measured phases and control signals from the experimental control system (Controller + PC) are used to produce RF drive tones for the AOMs. (b) The phase detector. A quarter-wave plate ($\lambda$/4), non-polarizing beam splitter (NPBS) and polarizing beam splitter (PBS) convert the quadratures of the input laser into intensities at the two output ports of the PBS. The ordinary heterodyne signal is produced at port H. Gold arrows indicate the beams' polarization.}
\end{figure}

A diagram of the optical setup can be found in Fig. \ref{fig:fig2}(a). We prepare five beams from the same narrow-linewidth 1064 nm fiber laser using a series of beam splitters. Each beam passes through an AOM, from which we collect the positive first order diffraction, such that each beam is upshifted by the AOM drive tone frequency. The five beams overlap in the center of a quartz vacuum cell, which will allow us to apply the quasiperiodic potential to a quantum gas in future experiments. Errors in the 72$^\circ$ relative angle of the lattice beams $\leq \pm 0.25^\circ$ should produce a change in the potential that is imperceptible to the quantum gas. Notably, an eight-fold quasiperiodic potential was shown to tolerate $\pm1^{\circ}$ deviations in beam alignment \cite{viebahn_matter-wave_2019}, which may empirically hold for the five-fold quasiperiodic potential. Details about the relative angle error and the alignment procedure can be found in Appendix \ref{app:angle}.

We independently adjust the intensity and frequency of each lattice beam by controlling the AOM's drive tone. The intensity noise of each beam is suppressed with stand-alone intensity control signals that are fed back to the AOM. On one beam (the ``reference'' beam) we use an electro-optic modulator (EOM) to apply an additional phase dither with modulation depth $\approx$ 1\% and modulation frequency of 80 MHz. We take samples from each beam using an uncoated glass surface and measure the phase of each beam relative to the reference beam. To minimize the accumulated phase error at the lattice location, we make this measurement as close as possible to the science cell, minimizing the propagation length between the lattice location and the phase measurement location, thus minimizing the potential for injected and uncompensated phase noise from the environment. Additionally, we use high-stability optics mounts and a vibrationally-isolated optical table for additional mechanical noise suppression.

Rather than using a simple interferometric phase measurement, which forfeits one quadrature of the phase information, we use the optical subassembly shown in Fig. \ref{fig:fig2}b, which preserves both quadratures of the phase information. More specifically, we circularly polarize one lattice beam before combining it with the other lattice beam on a non-polarizing beam splitter and then use a polarizing beam splitter rotated by 45$^\circ$ to project the light onto a rotated polarization basis. Both quadratures of the measured phase are thus imprinted onto the amplitude of the two output beams. More details about this assembly can be found in Appendix \ref{app:qpd}.

By controlling the frequency of each beam, we thus control the phase. The phase dither from the EOM allows us to detect and manipulate the change in optical power due to the changing optical phase in the radiofrequency (RF) regime, where there is less environmental noise. We measure the signals using eight (two per beam pair) home-built resonant RF photodetectors (Appendix \ref{app:rpd}) and feed the signal into a control circuit that adjusts the AOM drive tone frequencies to stabilize the optical phase. The control circuit processes the RF signals from the phase detector, along with control signals that determine the phase setpoint, to produce an error signal via an IQ mixing scheme (see Appendix \ref{app:pll} for more details). The error signal drives an active proportional-integral control servo, whose output controls the frequency of an RF tone that is generated by a voltage controlled oscillator. This tone passes through a variable gain amplifier before driving an AOM. When the phase loop is closed, the servo feeds back onto the frequency of the controlled lattice beam such that the phase difference between it and the reference beam follows the setpoint phase defined by the control signals. This capability allows us to program arbitrary phase evolution trajectories for each of the controlled beams, which in turn translates to arbitrary control over translation and phasonic configuration of the quasicrystal, subject only to the locking bandwidth of the phase control system.

\section{\label{sec:specs}System Characterization}

\begin{figure}[!t]
    \includegraphics[width=3in]{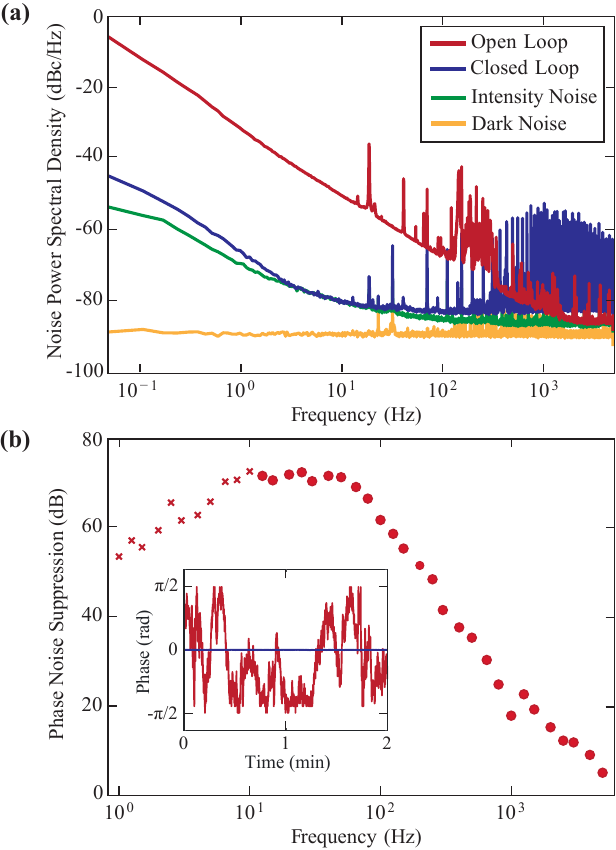}
    \caption{\label{fig:fig3}Phase noise suppression of the phase lock. (a) The phase noise power spectral density of one controlled lattice beam with the phase lock open (red) and closed (blue). The closed-loop measurement is dominated by residual fluctuations in the optical intensity of the beam (green). (b) The phase noise suppression measured with the dither scheme described in the main text. The filled circles are measured values, and the cross marks are lower bounds on the suppression. Note that the phase noise suppression appears to level out above 70 dB for frequencies below about 60 Hz, and the suppression remains positive for all measured frequencies $1~\mathrm{Hz} \leq f \leq  \mathrm{5~kHz}$. Time-series data (b, inset) demonstrate excellent optical phase stability when the phase lock is closed (blue) versus when it is open (red).}
\end{figure}

The performance of the phase control system was characterized in two ways: its ability to suppress environmental phase noise and the rate at which it can tune the phase of each laser. We made an out-of-loop measurement of the phase noise by recording the heterodyne signal from each of the four phase detectors shown in Fig.~\ref{fig:fig2}. The power spectral density of one record is shown in Fig.~\ref{fig:fig3}a. Compared with the open loop phase noise (red), closed loop operation shows substantial performance at frequencies $\lesssim$ 100 Hz (blue). The intrinsic noise of the photodetector is negligible. Most of the noise measured in the closed loop configuration can be attributed to the lasers' residual intensity noise (green). For frequencies $\gtrsim$ 300 Hz the closed loop spectrum contains a number of sharp features that seem to result from nonlinear interactions between the intensity and phase locks, and from the fiber laser's residual intensity noise. We note that these features' contribution to the variance of the phase is negligible.

We directly measure phase noise suppression after injecting a known amount of phase noise. A piezoelectric element introduces this noise by shaking one of the controlled beam's mirrors at a variable drive frequency. Comparing measurements of the shaking amplitude's Fourier component at the drive frequency, with the phase lock opened and closed, yields a measurement of the phase lock's noise suppression, as shown in Fig. \ref{fig:fig3}b. For frequencies $\lesssim$ 10 Hz, the residual closed-loop injected phase noise is still unresolvable below the intensity noise floor, giving a lower bound on the suppression factor. Typical phase noise suppression for our phase control system is $\gtrsim$ 70 dB for low frequencies. 

To measure the phase lock's speed, we ramp the phase setpoints of the controlled lattice beams at increasing frequencies until the phase lock fails. We observe that the measured phases of the beams closely follow the phase setpoints for modulation frequencies $\lesssim$ 350 kHz. This bandwidth is large enough to modulate the laser beam phases so as to accelerate the lattice to the lattice recoil velocity of lithium, which defines the relevant energy scales of the lattice dynamics we aim to explore in future experiments. The bandwidth of the control system is likely limited by the acoustic delay in the AOM.

\begin{figure}[!t]
\includegraphics[width=0.45\textwidth]{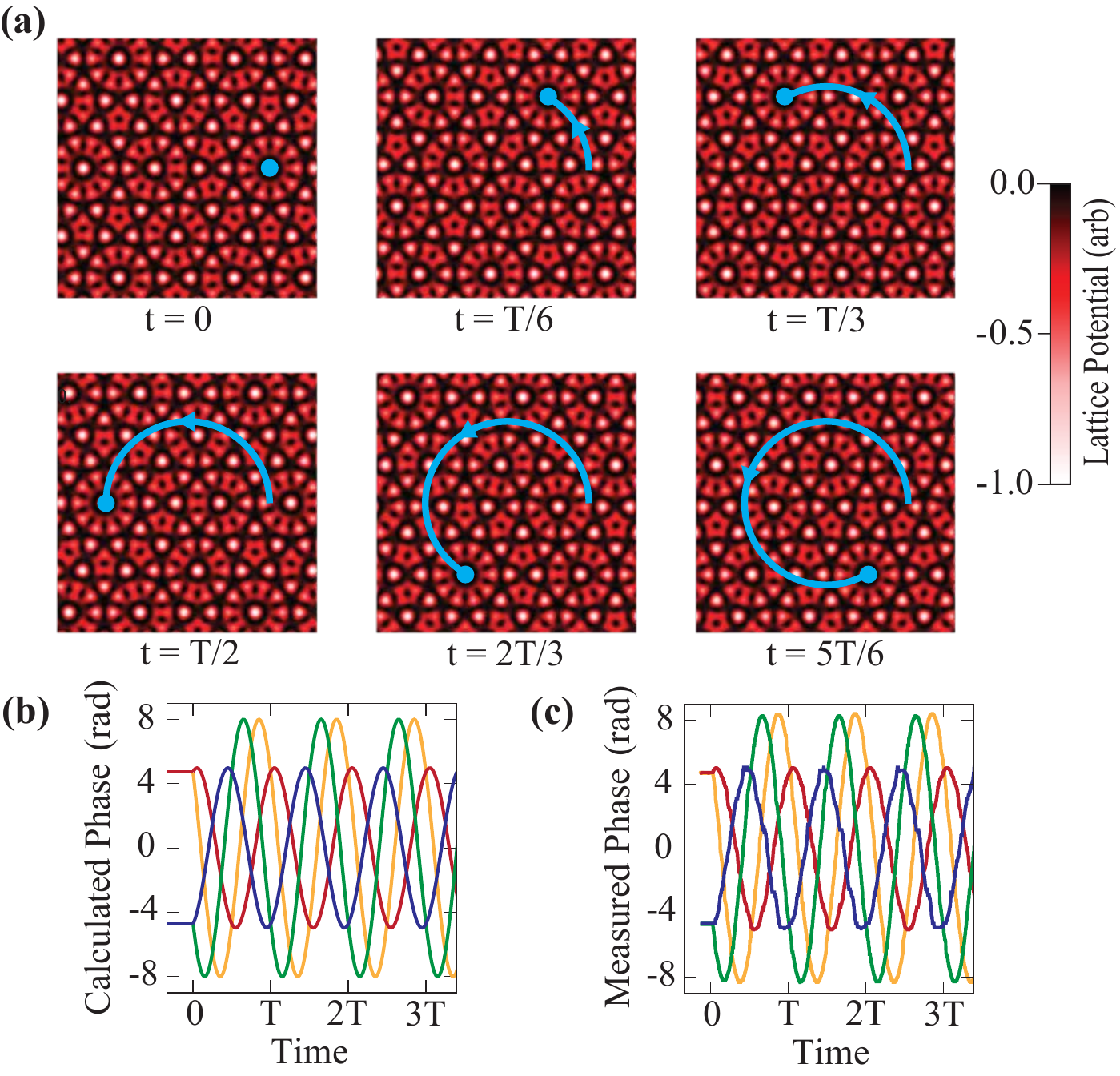}
\caption{\label{fig:fig4}Arbitrary translations of the quasicrystal lattice. (a) Calculated potential for six values of the lattice beam phases. In each panel, the phases are chosen so that the potential is simply translated (the blue circle shows the same lattice point in each panel). (b) Laser phases that cause the entire potential to be translated around a circle (the blue curve in (a)) over a period $T$. (c) Measured beam phases as a function of time. The control sequence measured in (c) corresponds to traversing the circle in (a,b) with a period $T\approx300~\mu$s, corresponding to lattice velocity 181 mm/s ($\approx$ 3.4 times the lattice recoil velocity for lithium).}
\end{figure}

\section{\label{sec:control}Phase Control}

By controlling the phases of four lattice beams relative to the reference beam, we have access to four degrees of freedom. Two degrees of freedom can be thought of as translations of the quasicrystal, which we explore in Section \ref{sec:trans}. The remaining degrees of freedom can be thought of as phasonic transformations of the quasicrystal, which we explore in Section \ref{sec:phason}. 

\begin{figure*}[!ht]
\includegraphics[width=0.9\textwidth]{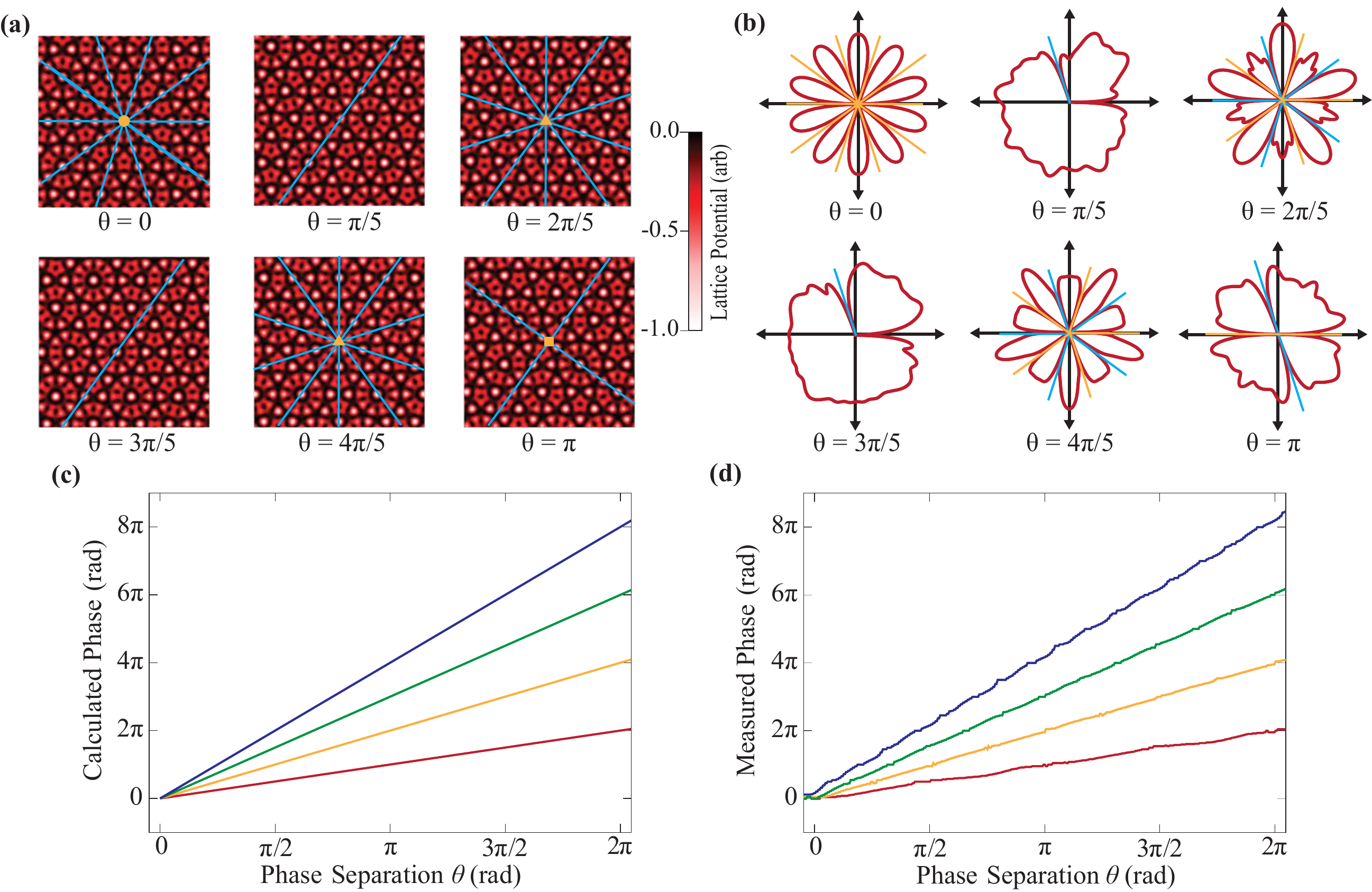}
\caption{\label{fig:fig5}Symmetry control of the quasicrystal lattice. (a) Calculated potential for six values of the phase separation $\theta$ between adjacent lattice beams. The local symmetries of the lattice (blue lines: reflection planes, yellow circles: $C_{10}$ rotation axes, yellow triangles: $C_5$ rotation axes, yellow squares: $C_2$ rotation axes) change over the evolution of the phase separation. When $\theta=\pi$, a $C_2$ rotation symmetry appears and creates a second axis of reflection orthogonal to the first. When $\theta$ is an integer multiple of $2\pi/5$, a $C_5$ rotation symmetry appears (see Appendix \ref{app:sym} for explanation). (b) Radial plots of the angular correlation (equation \ref{eqn:corr}) show zeros that indicate the existence of rotation (yellow lines) or reflection (blue lines) symmetry. (c) The phases calculated to generate changes of lattice geometry. Each panel in (a) is labeled with the corresponding value of $\theta$. (d) Measured beam phases as a function of $\theta$. In (c) and (d) the four different color traces are the phase trajectories of the four controlled lattice beams.}
\end{figure*}

\subsection{\label{sec:trans}Lattice Translation}

We achieve translations of the quasicrystal lattice by adjusting the optical phases of four of the lattice beams, relative to the fifth ``reference'' beam with wavevector $\mathbf{k}_5$. The relative phases $\varphi_j$ of the four controlled beams for a translation $\delta\mathbf{r}$ satisfy the relation 
\begin{equation}
\label{eqn:phase-tl}
    \varphi_j = - (\mathbf{k}_j-\mathbf{k}_5)\cdot\delta\mathbf{r},
\end{equation}
for laser wavevectors $\mathbf{k}_j$. Details about this relation and its derivation can be found in Appendix \ref{app:tl}. The translations are limited only by the speed of the phase control system, and are otherwise arbitrarily tunable. For example, we can cause the lattice to move along a circular path, as shown in Fig. \ref{fig:fig4}a by applying the phases shown in Fig. \ref{fig:fig4}b, which we calculate by feeding a circular trajectory into equation \ref{eqn:phase-tl}. Fig. \ref{fig:fig4}c shows the measured beam phases that demonstrate the lattice following a circular trajectory.

When the quasiperiodic lattice potential is applied to ultracold atoms, translations of the lattice in the laboratory reference frame are equivalent to translations of the atomic ensemble in the lattice frame. Our phase control scheme thus allows for nearly arbitrary control over the trajectory of the ensemble through the lattice. This level of control facilitates positioning the ensemble anywhere on the lattice, and imparting any momentum onto the ensemble, so long as the phase control signals are within the bandwidth of the phase lock. The 350 kHz bandwidth we demonstrate in this paper is large compared to the 95 kHz rate of phase evolution required to accelerate the atoms across the edge of the first pseudo-Brillouin zone, as defined in \cite{spurrier_semiclassical_2018}. This range of momentum control opens the door to direct probes of the geometry and topology of the effective energy band structure using schemes analogous to those used in \cite{brown_direct_2022}.

\subsection{\label{sec:phason}Phasonic Freedom}

The quasicrystal lattice supports two phasonic degrees of freedom, in addition to its two translational degrees of freedom. The lattice can be described in the ``cut-and-project'' picture as a two-dimensional slice through a five-dimensional cubic lattice. The two translational degrees of freedom correspond to motion along the projection plane, and the phasonic degrees of freedom correspond to motion orthogonal to the projection plane \cite{corcovilos_two-dimensional_2019,kosch_multifrequency_2022}. These degrees of freedom generate configurational changes in the quasicrystal \cite{corcovilos_two-dimensional_2019,steurer_quasicrystals_1997}, and can be used to dynamically change global symmetries in the lattice. Fig. \ref{fig:fig5}a shows a calculated plot of the lattice potential as the optical phases are varied as $\varphi_n=n\theta$ for integer $n$, which changes the phasonic degree of freedom (see Appendix \ref{app:sym}). In all six panels, there is one axis of reflection symmetry at $3\pi/5$, and only for certain phases do rotational symmetries appear, with a ten-fold ($C_{10}$) rotation symmetry appearing when all the phases are equal, a five-fold rotation symmetry ($C_5$) appearing when $\theta$ is an integer multiple of $2\pi/5$, and a two-fold rotation symmetry ($C_2$) appearing when $\theta = \pi$. These symmetries can be seen by plotting the quantity 
\begin{equation}\label{eqn:corr}
    K(\phi,0) = \int_0^\infty \mathrm{d}r \, \left(V(r,\phi) - V(r,0)\right)^2
\end{equation}
for lattice potential $V$ expressed in polar coordinates $(r,\phi)$. The zeroes of $K(\phi,0)$ indicate that the potential along a cut at angle $\phi$ is the same as the potential along the cut $\phi=0$. In practice, the integration is carried out numerically and out to very large distances $r\gg 1/|\mathbf{k}_j|$. Figs. \ref{fig:fig5}b show plots of $K(\phi,0)$, demonstrating that we can switch between $C_{10}$ symmetry, $C_5$ symmetry without $C_2$, or only $C_2$ symmetry. Fig. \ref{fig:fig5}c-d present calculated and measured phase trajectories that showcase our ability to dynamically change spatial symmetries of the lattice.

Recent results \cite{shimasaki_reversible_2024} demonstrate that periodic driving of the phasonic degree of freedom can be used to Floquet engineer the effective disorder strength in a one-dimensional quasicrystal. Our dynamic phase control scheme allows us to extend this method to two-dimensional quasicrystals, enabling directed experimental study of two-dimensional extensions of the Aubry-Andr\'{e} model, including probing the existence of metal/insulator phase transitions \cite{szabo_mixed_2020,strkalj_coexistence_2022}.

\section{Conclusion}

We have constructed a two-dimensional quasiperiodic optical lattice with programmable dynamic translational and phasonic control for use in a quantum simulator of quasicrystals.  We further find 350 kHz lock bandwidths and excellent phase noise suppression over experimentally relevant frequency scales. Such a degree of phase control brings several open questions about the dynamics of quasiperiodic systems within experimental reach.

Our phase control technique enables direct probes of quantum transport in quasiperiodic systems. Prior theoretical work \cite{spurrier_semiclassical_2018,sanchez-palencia_bose-einstein_2005} has established that quasicrystals support quasiperiodic analogs to Bloch oscillations in the shallow-lattice limit. Using the phase control system, we can move the atoms across the edge of the pseudo-Brillouin zone in the lattice frame, inducing quasiperiodic Bloch oscillations, which we can read out with real-space or Kapitsa-Dirac diffraction imaging. 

Additionally, the phase control system allows us to, in the shallow lattice limit, bring the atoms to points of high symmetry in the reciprocal lattice. This scheme should enable direct measurements of the anomalous group velocity induced by the topology of the quasicrystal.

The phasonic control accessible in our system opens the doors to an experimental realization of Thouless pumping \cite{corcovilos_two-dimensional_2019}. By driving periodic phasonic transformations of the quasicrystal and reading out the resulting spatial distribution of the atomic ensemble, we may be able to resolve quantized transport of the atoms. Beyond this, periodically driving the phasonic degrees of freedom could be used to Floquet engineer the strength of the disorder in the lattice \cite{shimasaki_reversible_2024}, enabling probes into the behavior of two-dimensional extensions of the Aubry-Andr\'{e} model.

\begin{acknowledgments}

We thank Tsz-Him Leung and Nathan Apfel for helpful discussions.

This work is based on work supported by the U.S. Air Force Office of Scientific Research under grant number FA9550-24-1-0229, by the U.S. National Science Foundation under grants PHY-2340760, DGE-2139841, and PHY-2402298, as well as by Yale University and the Yale Quantum Institute.\\

\end{acknowledgments}

\appendix

\section{\label{app:qpd}Quadrature Optical Phase Detection}
The phase control scheme presented in this paper involves separate control over both quadratures of the optical phase of each of the controlled beams. To accomplish this, we need to make measurements of both quadratures of the optical phase by sending the controlled beam and the reference beam into an interferometric phase detector. 

An ordinary Michaelson interferometer only provides information about one quadrature, since the two output signals are directly coupled. Instead, we extract information from both quadratures, by circularly polarizing one of the input beams, which results in a relative phase shift of $\pi/2$ between the horizontal and vertical components of the beam's polarization. We use a 50:50 non-polarizing beam splitter to combine the circularly polarized beam with a linearly polarized reference beam. We direct the combined beam into a polarizing beam splitter (PBS) mounted at 45$^\circ$, projecting the polarization state onto a rotated polarization basis. We thus produce beams at the outputs of the PBS whose intensities contain orthogonal phase quadratures. To illustrate this, consider two linearly polarized beams with amplitude $E_0$, angular frequency $\omega$, and polarization $\hat{\mathbf{x}}$, with one beam carrying an optical phase $\varphi$ relative to the other. The electric fields of the two beams are given by
\begin{align}
    \mathbf{E}_1 &= E_0 \mathrm{e}^{i\omega t + i\varphi} \hat{\mathbf{x}}\nonumber\\
    \mathbf{E}_2 &= E_0 \mathrm{e}^{i\omega t} \hat{\mathbf{x}}.
\end{align}
A quarter-wave plate in the path of one of the beams rotates its polarization to $(\hat{\mathbf{x}}+i\hat{\mathbf{y}})/\sqrt{2}$, and the 50:50 beam splitter combines the electric fields, producing a net electric field proportional to
\begin{equation}
    \mathbf{E}\propto E_0 \mathrm{e}^{i\omega t}\left( \mathrm{e}^{i\varphi} \hat{\mathbf{x}} + \frac{\hat{\mathbf{x}}+i\hat{\mathbf{y}}}{\sqrt{2}} \right).
\end{equation}
Projecting onto the rotated polarization basis $(\hat{\mathbf{x}}\pm\hat{\mathbf{y}})/\sqrt{2}$ with a PBS mounted at 45$^\circ$ gives output electric fields 
\begin{equation}
    \mathbf{E}_\pm = E_0  \mathrm{e}^{i\omega t} \left( \mathrm{e}^{i\varphi} + \frac{1\pm i}{\sqrt{2}} \right) \frac{\hat{\mathbf{x}}\pm\hat{\mathbf{y}}}{2}, 
\end{equation}
with intensities
\begin{equation}
    I_\pm\propto|\mathbf{E}_\pm|^2 = E_0^2 \left( 1 + \cos \left(\varphi\mp\frac{\pi}{4}\right) \right).
\end{equation}
$I_\pm$ both vary with $\varphi$, but are out of phase with each other by $\pi/2$ and therefore encode orthogonal quadratures of the optical phase. 

It is advantageous to push these phase signals into the RF regime to reduce $1/f$ environmental noise. We accomplish this by dithering the phase of one of the beams with modulation depth $\beta\sim1\%$ and modulation frequency $\Omega = 2\pi\times80$ MHz. The intensities of the two output beams of the phase detector then become
\begin{equation}\label{eqn:int-mod}
     I_\pm\propto E_0^2 \left( 1 + \cos \left(\varphi\mp\frac{\pi}{4} + \beta\sin\left(\Omega t\right)\right) \right).
\end{equation}
Since $\beta\ll1$, we can expand equation \ref{eqn:int-mod} in powers of $\beta$ and keep only up to first order terms, giving
\begin{equation}\label{eqn:int-qpd}
     I_\pm\propto E_0^2 \left( 1 - \beta\sin\left( \varphi\mp\frac{\pi}{4} \right)\sin\Omega t \right) + \mathcal{O}\left(\beta^2\right).
\end{equation}
The two beams' intensities therefore have, in addition to a constant term, a term oscillating at the modulation frequency. Both quadratures of the optical phase difference between the two beams are imprinted on the amplitudes of this oscillating term. By measuring the optical intensities of the two beams with a sufficiently fast photodetector, we can uniquely reconstruct the optical phase difference modulo $2\pi$.

\section{\label{app:rpd}Resonant Radiofrequency Photodetection}

\begin{figure}[!ht]
\includegraphics[width=0.25\textwidth]{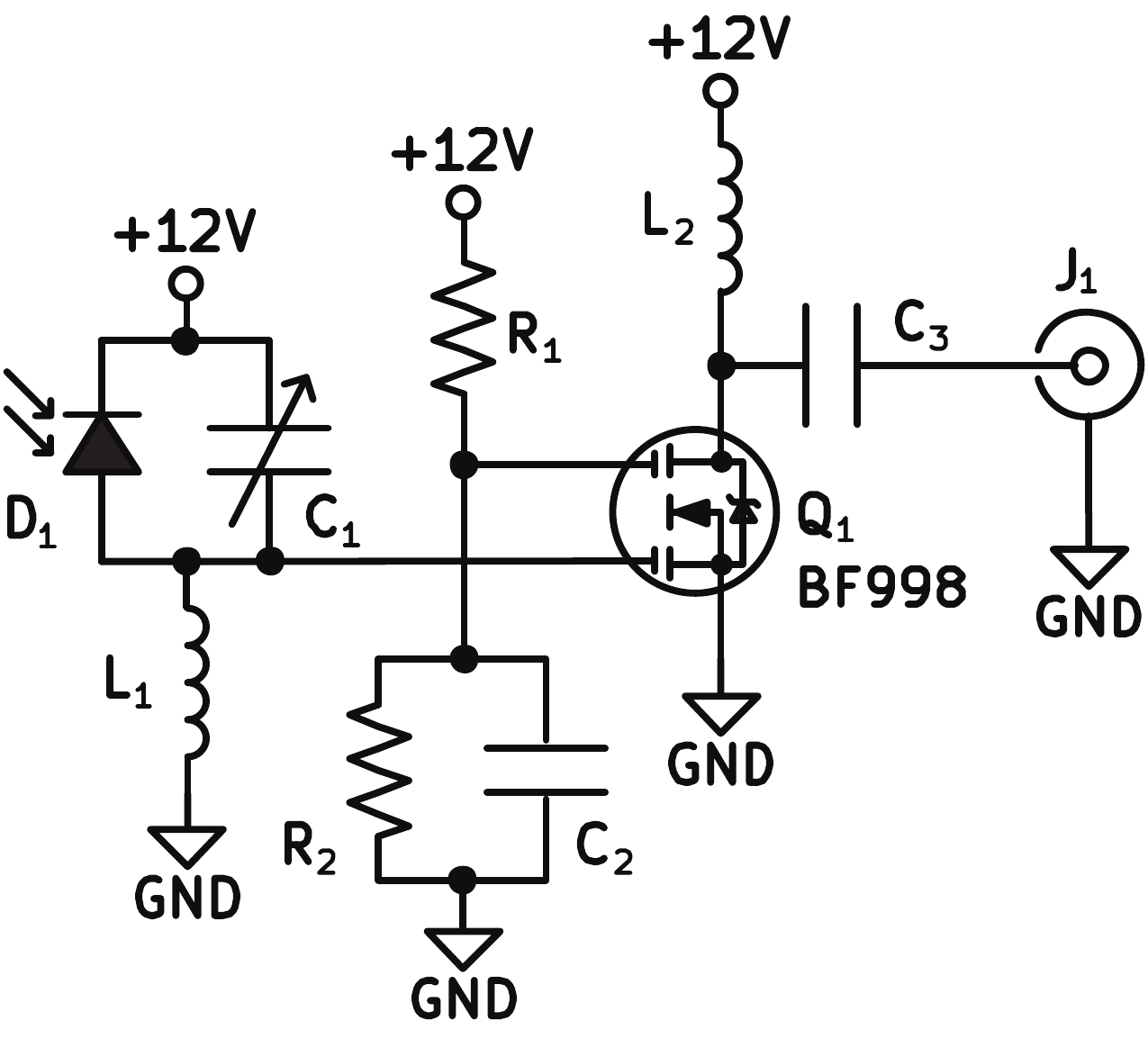}
\caption{\label{fig:figB1}The resonant-frequency photodetector circuit. The junction capacitance of photodiode $D_1$, trimmer capacitor $C_1$, and inductor $L_1$ form an $LC$ resonator that responds to optical signals at a certain tunable frequency. Signals from this are amplified by a common-source amplifier based on dual-gate MOSFET $Q_1$.}
\end{figure}

The phase information from the phase detection optics is encoded in an amplitude modulation at frequency $\Omega = 2\pi\times80$ MHz of two laser beams, as described in Appendix \ref{app:qpd}. We transduce this optical modulation into electrical signals with resonant-frequency photodetectors, a diagram of which can be found in Fig. \ref{fig:figB1}. The concept behind this circuit is that the inherent junction capacitance $C_\mathrm{J}$ of photodiode $D_1$ forms an LC resonator with inductor $L_1$. We add tunable capacitance $C_1$ in parallel to the photodiode so that we can tune the resonance frequency $\nu = 1/2\pi\sqrt{L_1(C_1+C_\mathrm{J})}$ of the detector. The quality factor of the resonance is limited by the intrinsic resistance $R$ of $D_1$, which gives the $LC$ resonator a bandwidth of $\gamma = R/2\pi L_1$. We choose the nominal value of $L_1=500$ nH, and tune trimmer capacitor $C_1$ such that the resonance is at $\nu=\Omega/2\pi=80$ MHz. The MTD3910W photodiode we use has $R\approx10$ $\Omega$, which gives the circuit a bandwidth $\gamma\approx3$ MHz. While the signal responsivity of the circuit could be increased by choosing a photodiode with smaller $R$, thereby increasing the quality factor of the $LC$ resonator, this would also decrease the bandwidth of the circuit, limiting the speed at which our phase control system can measure changes in optical phase. 

We amplify the signal from the $LC$ resonator with a common-source amplifier based on dual-gate MOSFET Q1 (BF998), which has a low noise figure and high bandwidth. One gate of Q1 accepts the signal from the $LC$ resonator, while the other gate is biased with voltage divider R1+R2 to set the gain of the amplifier, and capacitor C2 bypasses high frequency noise. Capacitor C3 AC couples the output of the circuit. Not pictured are impedance matching and power supply grooming components. 

\section{\label{app:pll}Phase Control Circuitry}

\begin{figure*}[!ht]
\includegraphics[width=0.9\textwidth]{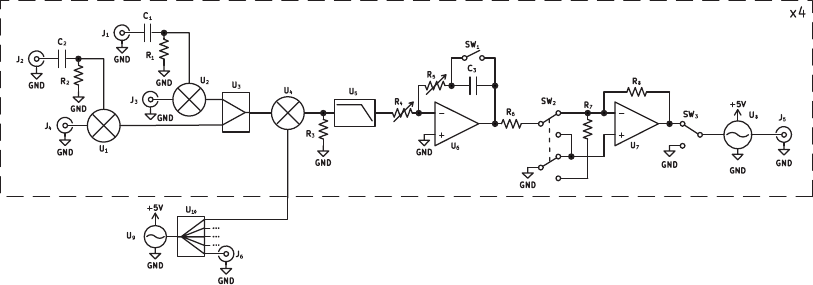}
\caption{\label{fig:figC1} The phase control circuit. The measured RF tones from the resonant photodiodes enter the circuit through coaxial connectors $J_1$ and $J_2$ and are AC-coupled by capacitors $C_1$ and $C_2$. Load resistors $R_1=R_2=50$ $\Omega$ set the input impedance of the circuit. Analog multipliers $U_1$ and $U_2$ multiply the RF tones by control signals that enter through coaxial connectors $J_3$ and $J_4$. The two resulting signals are combined on 0$^\circ$ splitter $U_3$ and demodulated at $\Omega$ by mixer $U_4$ and low-pass filter $U_5$, producing the error signal. A proportional-integral controller converts the error signal into an appropriate control signal. The proportional and integral gains of the controller can be adjusted by tuning potentiometers $R_4$ and $R_5$. The integral gain can be entirely turned off by closing switch $SW_1$ to bypass the feedback capacitor $C_3$. Switch $SW_2$, along with resistors $R_6$-$R_8$ and op amp $U_7$ form a unity gain amplifier that can be switched between inverting and non-inverting geometries, which allows us to change the sign of the open loop transfer function. Switch $SW_3$ opens/closes the feedback loop. In the closed-loop configuration, the control signal from the proportional-integral controller, after passing through the unity-gain amplifier, feeds into voltage-controlled oscillator $U_8$, controlling its oscillation frequency. The RF tone produced by $U_8$ exits the circuit through coaxial connector $J_5$. This entire assembly, demarcated by a dashed box, controls the phase of one of the four controlled lattice beams. Three more identical assemblies are also present, creating a total of four phase control channels. Voltage-controlled crystal oscillator $U_9$ produces the modulation tone $\Omega$ for the EOM phase dither, as well as the demodulation tone for each of the four channels. As such, the RF tone produced by $U_9$ is split by five-way 0$^\circ$ splitter $U_{10}$ into five equal portions. The four demodulation tone signals enter into their respective controller channel, and the phase dither tone exits the circuit through coaxial connector $J_6$. Auxiliary components like decoupling capacitors are not shown.}
\end{figure*}

The resonant photodetectors described in appendix \ref{app:rpd} convert the AC component of optical intensities $I_\pm$ (equation \ref{eqn:int-qpd}) into RF signals with amplitudes encoding two orthogonal quadratures of the optical phase $\varphi$, given by
\begin{equation}
    V_\pm\propto \sin\left(\varphi\mp\frac{\pi}{4}\right)\sin\Omega t.
\end{equation}
Scaling each of these signals by factor 
\begin{equation}\label{eqn:ctl}
    S_\pm = \mp\sin\left(-\theta\mp\frac{\pi}{4}\right)
\end{equation}
for angle setpoint $\theta$ and summing the two signals produces an RF signal with amplitude proportional to $\sin(\varphi-\theta)$, according to
\begin{equation}\label{eqn:err}
    V_+ S_+ + V_-S_- \propto \sin(\varphi-\theta)\sin\Omega t.
\end{equation}
Demodulating this signal at frequency $\Omega$ recovers just the amplitude, which we use as an error signal to lock the optical phase $\varphi$ to the setpoint phase $\theta$. 

A block diagram of the circuitry used to produce the signal in equation \ref{eqn:err} can be found in Fig. \ref{fig:figC1}. We accomplish the first multiplication step using Analog Devices AD835 analog multipliers. We then add the two signals with a 0$^\circ$ RF splitter (Mini-Circuits ADP-2-1+) before demodulating at $\Omega$ with a low LO-level mixer (Mini-Circuits ADE-1L) and an 8 MHz low-pass filter (Mini-Circuits SXLP-8+). The driving tone on the LO port of the mixer comes from the same crystal oscillator (Crystek CVSS-945-80.000) that we use to drive the EOM phase dither, ensuring that the modulation and demodulation tones are exactly matched. 

The demodulated signal functions as an error signal. We use this error signal alongside an active proportional-integral (PI) controller with adjustable gains to feed back onto the optical phase by adjusting the output frequency of the voltage-controlled oscillator (Mini-Circuits ROS-95-419+) that drives the AOM. The frequency shifts applied to the beam this way stabilize the phase by advancing the optical phase (which is the integral of the optical frequency over time) towards the zero of the error signal. Since the error signal is proportional to $\sin(\varphi-\theta)$, the control loop enforces $\varphi=\theta + n\pi$ for integer $n$. Since the slope of the error signal against $\varphi-\theta$ bears opposite sign for even and odd $n$, only one parity of $n$ corresponds to stable locking points, meaning we can uniquely define $\varphi$ modulo $2\pi$ for a given setpoint $\theta$. By making dynamic changes to $\theta$ by changing the signals $S_\pm$ in time, we accomplish dynamic control of the optical phase $\varphi$.

\section{\label{app:tl}Quasicrystal Lattice Translation}

We can, by making the phases of the lattice beams follow certain trajectories, cause the lattice to translate. To calculate the required phases, we begin by writing the lattice potential as 
\begin{equation}
    V = \frac{V_0}{2}\left| \sum_{m=1}^5 \mathrm{e}^{i\mathbf{k}_m\cdot\mathbf{r}+i\varphi_m} \right|^2,
\end{equation}
where $V_0$ is the lattice depth, and $\mathbf{k}_m$ and $\varphi_m$ is the wavevector and phase, respectively, of the $m$-th lattice beam. The square modulus can be expanded as
\begin{align}
    V&=\frac{V_0}{2}\sum_{m=1}^5\sum_{n=1}^5\mathrm{e}^{i\left(\mathbf{k}_m-\mathbf{k}_n\right)\cdot\mathbf{r}+i\left(\varphi_m-\varphi_n\right)}\\
    &= \frac{5V_0}{2} + V_0\sum_{m=1}^5\sum_{n<m}\cos\big(\left(\mathbf{k}_m-\mathbf{k}_n\right)\cdot\mathbf{r}+\left(\varphi_m-\varphi_n\right)\big),\nonumber\label{eqn:pot}
\end{align}
highlighting that the lattice is composed of ten (5 choose 2) standing waves formed from the interference of each unique pair of the five laser beams. As such, the phases of the beams only enter the potential as differences, allowing us to, without loss of generality, define one beam to have $\varphi_5 = 0$, and define the other four phases relative to this one. This scheme maps well onto the experimental setup, where we measure and control the phases of four ``controlled'' beams relative to that of the fifth ``reference'' beam.

To accomplish a translation, we apply a phase shift $\delta\varphi_m$ to each of the four controlled beams such that the potential appears to have been shifted by displacement $\delta\mathrm{r}$. The summand of equation \ref{eqn:pot} then must be transformed such that 
\begin{align}
    &\cos\big(\left(\mathbf{k}_m-\mathbf{k}_n\right)\cdot(\mathbf{r}+\delta\mathrm{r})+\left(\varphi_m-\varphi_n\right)+\left(\delta\varphi_m-\delta\varphi_n\right)\big)\nonumber \\ &= 
    \cos\big(\left(\mathbf{k}_m-\mathbf{k}_n\right)\cdot\mathbf{r}+\left(\varphi_m-\varphi_n\right)\big).
\end{align}
This is satisfied if, for all $(m,n)$, 
\begin{equation}
    \delta\varphi_m - \delta\varphi_n = -\left( \mathbf{k}_m - \mathbf{k}_n \right)\cdot\delta\mathbf{r}.
\end{equation}
It is sufficient to define $\delta\varphi_m$ for all $m$ relative to $\delta\varphi_5 = 0$ as
\begin{equation}
    \delta\varphi_m = -\left( \mathbf{k}_m-\mathbf{k}_5 \right)\cdot\delta\mathbf{r},
\end{equation}
recovering equation \ref{eqn:phase-tl}. This enables us to cause the lattice to translate by any arbitrary displacement by applying the correct phase shifts to the four controlled beams, allowing us to set the lattice to move along arbitrary trajectories with arbitrary velocities, so long as our phase control system is fast enough.

\section{\label{app:sym}Phasonic Control of Lattice Symmetries}

Beyond lattice translations, we can use the phase control system presented in this work to achieve phasonic transformations of the lattice. In the cut-and-project picture \cite{corcovilos_two-dimensional_2019,kosch_multifrequency_2022}, these transformations correspond to translation of the projection window orthogonal to the cut plane. Using this degree of freedom, we can dynamically change the global symmetries of the lattice. As an example, suppose that we are interested in lattices with a $C_5$ rotation symmetry. This means that the potential, as defined in equation \ref{eqn:pot}, must be invariant under rotation by $2\pi/5$, i.e.
\begin{align}
\label{eqn:c5}
    V(\mathbf{r}) &= V\left(R_{2\pi/5}\mathbf{r}\right),
\end{align}
where $R_\theta$ is the matrix that represents rotations by angle $\theta$ in $2\mathrm{D}$. The potential can, as discussed in Appendix \ref{app:tl}, be written as a sum over ten standing waves, and equating these terms on both sides of equation \ref{eqn:c5} demands that, in order for $C_5$ symmetry to be preserved,
\begin{align}\label{eqn:c5-2}
    &\cos\big(\left(\mathbf{k}_m-\mathbf{k}_n\right)\cdot \mathbf{r}+\left(\varphi_m-\varphi_n\right)\big) \nonumber\\ &=\cos\big(\left(\mathbf{k}_m-\mathbf{k}_n\right)\cdot R_{2\pi/5}\mathbf{r}+\left(\varphi_m-\varphi_n\right)\big).
\end{align}
The rotation matrix can be thought of as rotating $\mathbf{r}$ by angle $2\pi/5$ or rotating $(\mathbf{k}_m-\mathbf{k}_n)$ by angle $-2\pi/5$. Recalling that, as defined in Fig. \ref{fig:fig1}(b), the lattice wavevectors satisfy the cyclic rotation relationship
\begin{equation}
    \mathbf{k}_{m+1} = R_{2\pi/5}\mathbf{k}_m,
\end{equation}
where $\mathbf{k}_6$ is taken to be $\mathbf{k}_1$. This means the condition defined in equation \ref{eqn:c5-2} can be rewritten as 
\begin{align}
    &\cos\big(\left(\mathbf{k}_m-\mathbf{k}_n\right)\cdot \mathbf{r}+\left(\varphi_m-\varphi_n\right)\big) \nonumber\\ &=\cos\big(\left(\mathbf{k}_{m-1}-\mathbf{k}_{n-1}\right)\cdot\mathbf{r}+\left(\varphi_m-\varphi_n\right)\big).
\end{align}
This condition is satisfied if each nearest-neighbor pair of lattice beams $m,n$ has the same phase difference $\varphi_m-\varphi_n\equiv\theta$, implying also that each next-nearest-neighbor pair of lattice beams has phase difference $2\theta$. This condition is satisfied when $\theta=2N\pi/5$ for integer $N$, because of the $2\pi$ periodicity of the cosine function.

We use a similar line of reasoning to find the condition that allows for $C_2$ symmetry in the lattice potential. In this case, the potential must satisfy
\begin{equation}\label{eqn:c2}
    V(\mathbf{r}) = V\left(R_\pi\mathbf{r}\right) = V(-\mathbf{r}),
\end{equation}
which implies that 
\begin{align}
     &\cos\big(\left(\mathbf{k}_m-\mathbf{k}_n\right)\cdot \mathbf{r}+\left(\varphi_m-\varphi_n\right)\big) \nonumber\\ &=\cos\big(-\left(\mathbf{k}_{m}-\mathbf{k}_{n}\right)\cdot\mathbf{r}+\left(\varphi_m-\varphi_n\right)\big).
\end{align}
We can separate these cosine terms into components that are odd in $\mathbf{r}$ and those that are even in $\mathbf{r}$ as 
\begin{align}
     &\cos\big(\pm\left(\mathbf{k}_m-\mathbf{k}_n\right)\cdot \mathbf{r}\big)\cos\left(\varphi_m-\varphi_n\right) \nonumber\\&-\sin\big(\pm\left(\mathbf{k}_m-\mathbf{k}_n\right)\cdot \mathbf{r}\big)\sin\left(\varphi_m-\varphi_n\right).
\end{align}
To satisfy equation \ref{eqn:c2}, we need the component that is odd in $\mathbf{r}$ to be zero, which is satisfied when the phase difference $\varphi_m-\varphi_n$ between any two beams is an integer multiple of $\pi$.

Consider the case where the lattice beam phases are given as $\varphi_n = n\theta$. When $\theta=2N\pi$ for integer $N$, all five phases are zero (modulo $2\pi$), which satisfies both the $C_2$ and the $C_5$ condition. This results in a lattice with a global $C_{10}$ symmetry. When $\theta=2N\pi/5$ for integers $N$ that are not integer multiples of 5, we preserve the $C_5$ symmetry, but break the $C_2$ symmetry. When $\theta=N\pi$ for odd $N$, the $C_2$ symmetry remains while the $C_5$ symmetry disappears. This can be seen in Fig. \ref{fig:fig5} in the body of the text.

It is worth noting that for the different potentials formed for different values of $\theta$, while they have different symmetries, the energy band spectrum is the same. This is because any shifting of the optical phases induces a strictly unitary transformation on the Hamiltonian, which,  in the shallow lattice limit, can be interpreted as a gauge transformation on the wave functions, necessarily leaving its spectrum invariant. While surprising, this invariance can be understood as follows: the four phase degrees of freedom correspond to, in the cut-and-project picture, translations of the five-dimensional cubic lattice, which do not affect the global properties of the lattice.

\section{\label{app:angle}Angular Alignment of Lattice Beams}

Deviations in the angular alignment of the lattice beams from the perfect 72$^\circ$ case, e.g. from spatial misalignment, can affect the quality of the quasiperiodic lattice by limiting the long-range order and breaking global symmetries. Suppose that one of the lattice beams with intended wavevector $\mathbf{k}_j$ is tipped by angle $\delta$ relative to the perfect 72$^\circ$ alignment such that it has wavevector
\begin{equation}
    \tilde{\mathbf{k}}_j = R_\delta\mathbf{k}_j.
\end{equation}
For small $\delta$, the rotation matrix $R_\delta$ (see Appendix \ref{app:sym}) can be approximated such that
\begin{equation}
    \tilde{\mathbf{k}}_j = \mathbf{k}_j+\delta R_{\pi/2}\mathbf{k}_j + \mathcal{O}(\delta^2).
\end{equation}
This modified wavevector enters into the optical lattice potential as the change
\begin{equation}
    \mathrm{e}^{i\mathbf{k}_j\cdot\mathbf{r}}\to\mathrm{e}^{i\tilde{\mathbf{k}}_j\cdot\mathbf{r}} = \mathrm{e}^{i\delta (R_{\pi/2}\mathbf{k}_j)\cdot\mathbf{r}}\mathrm{e}^{i\mathbf{k}_j\cdot\mathbf{r}}.
\end{equation}
For the potential to appear unaffected by the angular deviation, we require the first exponential on the right-hand side to be close to unity. This happens when
\begin{equation}
    \left|\delta (R_{\pi/2}\mathbf{k}_j)\cdot\mathbf{r}\right|\ll1,
\end{equation}
which holds when 
\begin{equation}
    |\mathbf{r}|\ll L\equiv\frac{1}{\delta|\mathbf{k}|}.
\end{equation}
A stringent requirement on the beam angle accuracy is to enforce $L$ to be larger than the beam waist radius $w$ of the lattice beams. This requirement demands the relative angle of the lattice beams to be aligned within $72\pm$0.25$^\circ$. This demand ensures that the length scale over which the lattice has the desired symmetries is larger than the actual size of the lattice constrained by the finite lattice beam diameter.

In practice, the required alignment accuracy will be achieved by using atoms trapped in the lattice to perform iterative experiments that cycle between beam alignment adjustments and matter wave diffraction from all 1D periodic optical lattices created by only pairs of nearest-neighbor lattice beams.

\bibliography{bib}

\end{document}